\documentclass[a4paper,11pt]{article}
\pdfoutput=1 

\usepackage{jheppub} 

\usepackage[T1]{fontenc} 

\def\be{\begin{equation}}
\def\ee{\end{equation}}
\def\ba{\begin{eqnarray}}
\def\ea{\end{eqnarray}}
\def\no{\nonumber}

\title{Phase transition and entropy spectrum of BTZ black hole in three-dimensional gravity with torsion}

\author[a,b,1]{Meng-Sen Ma,\note{Corresponding author.}}
\author[a]{Ren Zhao}

\affiliation[a]{Institute of theoretical physics, Shanxi Datong University, 037009
Datong, China}
\affiliation[b]{Department of Physics, Shanxi Datong University, 037009 Datong, China}

\emailAdd{mengsenma@gmail.com}
\emailAdd{zhao2969@sina.com}

\abstract{In this paper, we study the phase transition and the entropy spectrum of BTZ black hole obtained
in a model of three-dimensional gravity with torsion. By calculating the heat capacity we find that the BTZ black hole we considered will
experience phase transition at some critical point. This indicates that the critical
behaviors of black holes do not only depend on the spacetime metric, but have to do with the theory of gravity under consideration.
In addition we derived the entropy spectrum of the BTZ black hole according to the quasinormal modes(QNMs) and the adiabatic invariance. It shows that
 the area or entropy spectrum will also rely on the concrete gravitational action.
}

\begin{document}
\maketitle
\flushbottom

\section{Introduction}

  Black holes as a prediction of general relativity(GR) have played a great role in the
 development of modern theoretical physics. Specially the discovery of Hawking radiation\cite{hawking1} makes black hole physics
to be a subject at the intersection of theories of gravity, quantum mechanics, statistical physics and field
theory. Since then the thermodynamics of black holes has received a lot of attention and the thermodynamic properties of
 many black holes have been studied extensively\cite{York,Wald1,Wald2,Strominger,Nojiri}. Among them the BTZ black hole in three dimensional gravity is of particular importance\cite{BTZ}.
 In lower spacetime many physical problems can be quite simpler.

 In general relativity and many other theories of gravity
 curvature plays an essential role , while torsion is received
less attention. However, torsion has also its geometrical meaning and plays some role
in gravitation theory. Since 1970s many theories of gravity with torsion have been proposed, such as Poincar\'{e} gauge gravity, de Sitter gauge gravity,
teleparalell gravity, $f(T)$ gravity et.al. In particular Mielke and Baekler proposed a model of three dimensional gravity with torsion (MB model),
 which also has the BTZ black hole as solution\cite{MB1,MB2}. This model aroused the following researches on the thermodynamics of BTZ black hole with torsion and
 AdS/CFT with torsion\cite{Hehl, Blagojevic1,Blagojevic2}.

 Phase transition and critical phenomenon are important characteristics of the
 ordinary thermodynamics. Thus the natural question to ask is whether there also exists the phase transition in the black hole thermodynamics.
Because black holes also have the standard thermodynamic
quantities, such as temperature, entropy, even the pressure and volume\cite{Dolan,Cvetic,RBM}, their critical phenomena may be similar
 to ones in the ordinary thermodynamic system.
 Until now there are mainly two different views about the phase transition of black holes in equilibrium.
 The first one, which is originated from Davies\cite{Davies}, insists that the phase transition of black holes turn up at the point where
 the heat capacities diverge. This characteristic is also present in ordinary thermodynamic systems which
  exhibit second order phase transitions. Many researches on this topic have been carried out\cite{Hut,Mazur,Lousto,Wu}.
  In this sense there is no phase transition for BTZ black hole obtained in GR.
Because the heat capacity of the ordinary BTZ black hole is always positive.
  Recently, based on Hawking-Page phase transition\cite{hawking2} some interesting works on asymptotically anti-de Sitter black hole have been done,
   which show that there exists phase transition similar to the van der Waals-Maxwell
vapor-liquid phase transition\cite{chamblin,chamblin1,Lemos, RBM,Tian,RBM2}.  Another point of view thinks the phase transition takes place
in the extremal limit of black holes. According to the fluctuation theory of equilibrium thermodynamics, some second moments
of relevant quantities for Schwarzschild black holes, Kerr black holes, RN black holes and
(charged) dilaton black holes have been calculated\cite{Pavon1,Pavon2,Cai1,Cai2}. It shows that nothing special happens for them at the Davies's points. But the second moments will
diverge when non-extremal black holes become extremal one. Moreover this kind of phase transition
 may be relevant to the quasinormal modes of black holes\cite{Wang1,Wang2,Jing,Wang3}.

In this paper we first study the phase transition and critical behavior
of BTZ black hole obtained in the MB model. Although the BTZ solution is the same as the one
obtained in GR, the different actions will make their thermodynamics very different. The modified action in the MB model
will correct the conserved charges such as mass and angular momentum. Correspondingly the entropy of BTZ black hole and the first
law of black hole thermodynamics will also been changed. Based on these results one can reanalyze the phase transitions of
the BTZ black hole  in the two senses mentioned above.  It shows that
 the heat capacity of the BTZ black hole in the MB model is not always positive any more, but changes signs at some points and may diverge at the critical point.
 Thus for the BTZ black hole in the MB model phase transition exists. Next, according to the fluctuation theory of thermodynamics we recalculate
 some second moments. They are still finite in the non-extremal case and diverge in the extremal case. This indicates that the second viewpoint which
 insists the phase transition takes place in the extremal limits of black holes still holds.  Moreover, according to the QNMs and the adiabatic invariance
 we studied the area spectrum and entropy spectrum of
 the non-rotating BTZ black hole  and rotating BTZ black hole in the MB model. Because the QNMs are independent of torsion,
 the area spectrum obtained is also irrelevant to torsion explicitely.

The paper is arranged as follows: in the next section we simply
introduce the MB model and its BTZ black hole solution and the corresponding thermodynamic quantities. In
section 3  we will calculate the heat capacity at constant angular momentum and some second moments, by which we
 discuss the phase structure and critical
phenomena of the BTZ black holes. In section 4 the area spectrum and entropy spectrum of the BTZ black hole are given.
We will make some concluding remarks in
section 5.

\section{Mielke-Baekler model and  thermodynamic quantities of the BTZ black hole}

To express theories of gravity one can employ
two dynamically independent one-forms: the connection $\omega^a_{~b}$ and vielbein $e^a$ . In GR
 these two fields are assumed to be linked by
the torsion-free condition $T=0$.
 We can define curvature and torsion 2-forms out of $\omega^a_{~b}$ and vielbein $e^a$ by

 \ba
 T^a&=&de^a+\omega^a_{~b}\wedge e^b \\
 R^a_{~b}&=&d\omega^a_{~b}+\omega^a_{~c}\wedge \omega^c_{~b}
 \ea
In geometry, $\omega^a_{~b}$
and $e^a$ reflect the affine and metric properties of the manifold, while in physics,
torsion and curvature may be related to energy momentum tensor and spin current respectively.

Now we should introduce the topological three dimensional gravity model with torsion proposed by Mielke and Baekler\cite{MB1,MB2}, which is a natural
generalization of Riemannian GR with a cosmological constant.
The  gravitational action is usually written as

\be\label{action}
I=\int 2ae^{a}\wedge R_{a}-\frac{\Lambda }{3}\epsilon _{abc}e^{a}\wedge e^{b}\wedge e^{c}+\alpha
_{3}\left( \omega ^{a}\wedge d\omega _{a}+\frac{1}{3}\varepsilon _{abc}\omega
^{a}\wedge\omega ^{b}\wedge\omega ^{c}\right) +\alpha _{4}e^{a}\wedge T_{a}~,
\ee
where the dual expression $R_a$ and $\omega_a$ are defined by $R^{ab}=\epsilon^{abc}R_c$ and $\omega^{ab}=\epsilon^{abc}\omega_c$. In Eq.(\ref{action}) the first term corresponds to the Einstein-Cartan action with $a=\frac{1}{16\pi G}$. The second term is the cosmological term which is a constant. The last
two terms are the Chern-Simons term and the Nieh-Yan term which should be played particular stress.

Using curvature we can construct characteristic classes, integrations of which
will be topological invariants of the manifold, which reflect the global properties of the manifold.
The well known characteristic classes for a four dimensional real manifold
are Pontrjagin and Euler classes\cite{Nakahara}

\ba
P&=&\frac{1}{8\pi^2}R^{ab}\wedge R_{ab}\\
E&=&\frac{1}{32\pi^2}\varepsilon_{abcd}R^{ab}\wedge R^{cd}
\ea
The integration of $P$ and $E$ on the manifold take integer number, which distinguish topologically
different manifolds.

It is well known that locally
\ba
P_4 &=& dQ, \no\\
Q &=& \omega \wedge d\omega + \frac{2}{3}\omega\wedge \omega\wedge \omega
\ea
is the local
Chern-Simons (CS) form, which has many applications in physics.
One of the important feature of CS modified gravity is its
emergence within predictive frameworks of more fundamental theories. For
example, the low energy limit of string theory comprises general relativity
with a parity violating correction term, that is nothing but the Pontryagin
density. This term is crucial for cancelling gravitational anomaly in string
theory through Green-Schwartz mechanism.

The Nieh-Yan(N-Y) form is a special 2-form only for the Einstein-Cartan manifold\cite{NY,Zanelli}.
On four dimensional manifold $M$ it can be written as
\ba
&&N=T^a\wedge T_a+R_{ab}\wedge e^a\wedge e^b=dQ_{NY}, \no\\
&&Q_{NY}=e^a\wedge T_a
\ea
N-Y form is a kind of Chern-Simons form and will have its application
to manifold with boundary and reflect the role of torsion in geometry.

After variation to the connection $\omega^a_{~b}$ and vielbein $e^a$, two vacuum equations can be obtained from the MB action
\ba
T^a&=&\frac{p}{2}\varepsilon^a_{~bc}e^b\wedge e^c, \\
R^a&=&\frac{q}{2}\varepsilon^a_{~bc}e^b\wedge e^c
\ea
with the two constant coefficients $p, q$ defined by
$p=\frac{\alpha _{3}\Lambda +\alpha _{4}a}{\alpha
_{3}\alpha _{4}-a^{2}}$ and $q=-\frac{\alpha _{4}^{2}+a\Lambda }{\alpha _{3}\alpha
_{4}-a^{2}}$.

The curvatures in Einstein-Cartan geometry can be connected to its counterpart in Riemannian geometry.
In particular in three dimensional spacetime the equations above can be simplified to  equations without torsion
\be\label{notorsion}
\tilde{R}^a=\frac{\Lambda_{eff}}{2}\varepsilon^a_{~bc}e^b\wedge e^c
\ee
where $\tilde{R}^a$ is the  curvature without torsion and $\Lambda_{eff}=q-\frac{1}{4}%
p^{2}$ is the effective cosmological constant. One can let $\Lambda _{eff}=-\frac{1}{l^{2}}<0$ to construct an asymptotically anti-de Sitter space.

As in the three dimensional Einstein equation, Eq.(\ref{notorsion}) has the well-known BTZ solution. But in this case torsion is contained in the gravitational action.
 The metric is
\be
ds^{2}=-N\left( r\right) ^{2}dt^{2}+\frac{1}{N\left( r\right) ^{2}}%
dr^{2}+r^{2}\left( d\phi +N_{\phi }dt\right) ^{2}
\ee
where
\begin{equation}
N\left( r\right) ^{2}=\frac{r^{2}}{l^{2}}-M_0+\frac{J_0^{2}}{4r^{2}},\text{
\ }N_{\phi }\left( r\right) =\frac{J_0}{2r^{2}}~.
\end{equation}%
Here we have considered $8G=1$. This metric is the same as the one in GR except the $l=1/\sqrt{-\Lambda _{eff}}$ here. For this metric there are two
horizons, the outer one $r_+$ and the inner one $r_{-}$.
From $N^2(r)=0$, one can obtain the expressions of the both horizons:
\be\label{horizon}
r^2_{\pm}=\frac{M_0l^2}{2}(1\pm \Delta), \quad \Delta=[1-(J_0/M_0l)^2]^{1/2}
\ee
Conversely, $M_0$ and $J_0$ can be expressed as follows: \be
M_0=\frac{r_{+}^2+r_{-}^2}{l^2}, \quad J_0=\frac{2r_+r_{-}}{l} \ee

Because Hawking radiation is just a kinematic effect, which only
depends on the event horizon and is irrelevant to the dynamical
equations and the gravitational theories. Therefore the temperature
of BTZ black hole in the MB model has the similar form as in GR,
which is
\be \label{temp}
T=\frac{r_{+}^2-r_{-}^2}{2\pi l^2r_{+}}
\ee
Certainly because of the existence of $l$, the temperature is relevant to the coefficients $\alpha_3,\alpha_4, \Lambda$ of MB Lagrangian.
Define
\be
\Omega_H=-\frac{g_{t\phi}}{g_{\phi\phi}}|_{r_{+}}=\frac{J_0}{2r_+^2}
\ee
which can be regarded as the angular velocity of BTZ black hole.
The entropy of BTZ black hole in GR can be obtained by means of Euclidean action method, which is
\be\label{BTZentropy}
S_0=4\pi r_+
\ee
which is twice the perimeter of the horizon. In fact the entropy can also be derived by employing the first law of
black hole thermodynamics. Because black holes are also thermodynamic systems, they must obey the first law of thermodynamics.
For the BTZ black hole in GR it is
\be
dM_0=TdS_0+\Omega_{H} dJ_0
\ee
Thus the entropy can be derived according to $T$ and $M_0$, namely
\be\label{calentropy}
S_0=\int T^{-1}dM_0+C_0
=\int T^{-1}(\frac{\partial M_0}{\partial ~r_{+}})|_{J_0}dr_{+}+C_0
\ee
It should be noted that the $J_0$ in the integration should be taken as constant. One can express $M_0$ and $T$ as function of $r_{+}$ and $J_0$, which are
\ba
M_0=\frac{r_{+}^2}{l^2}+\frac{J_{0}^2}{4r_{+}^2},\quad
T=\frac{r_{+}^2-\frac{l^2J_0^2}{4r_{+}^2}}{2\pi l^2r_{+}}
\ea
Substituting them into Eq.(\ref{calentropy}), one can easily obtain the result in Eq.(\ref{BTZentropy}).

For the MB model we can also employ this method to derive the entropy of BTZ black hole with torsion. But things are slightly different.
 Blagojevic et.al have proved that the gravitational conserved charges in the MB model should be\cite{Blagojevic3}
\be\label{charge}
M=M_0+2\pi\alpha_3(\frac{pM_0}{2}-\frac{J_0}{l^2}), \quad J=J_0+2\pi\alpha_3(\frac{p J_0}{2}-M_0)
\ee
Obviously when $\alpha_3=0$ they will return to their conventional interpretation as energy and angular
momentum, as with the BTZ metric in general relativity. In this case, if the first law of black hole thermodynamics still holds, the relation should be
\be\label{firstlaw}
dM=TdS+\Omega_{H} dJ
\ee
Therefore the $M$ and $J$ should be used in the Eq.(\ref{calentropy}),
\be
S=\int T^{-1}dM+C
=\int T^{-1}(\frac{\partial M}{\partial ~r_{+}})|_{J}dr_{+}+C
\ee
One thing to note is that in the integral $r_{-}$ cannot be regarded as constant but depends on $r_+$ and $J$. The entropy can be obtained easily
\be\label{BTZentropytor}
S=4\pi r_{+}+4\pi^2\alpha_3(pr_{+}-\frac{2r_{-}}{l})
\ee
This result agrees with the one Blagojevic et.al obtained using the partition function with the aid of  Euclidean action method\cite{Blagojevic1}.
It differs from the Bekenstein-Hawking result\cite{JDB,BCH} by an additional term
coincides with Solodukhin's result if $p=0$\cite{SN}. Black hole entropy is not always equated with one-quarter the
event horizon area. In fact it is related to the gravitational theory under consideration.  It can be easily verified  that
in the MB model the entropy, temperature and the conserved charges not only satisfy the first law of thermodynamics but also fulfill
Smarr-like formula
\be
M=\frac{1}{2}TS+\Omega_{H}J
\ee
This further implicates that with torsion the BTZ black hole can still seemed as a thermodynamic system and the thermodynamic laws still hold.
It should be noted that in the expression of entropy of BTZ black hole  no torsion exists explicitly, only $\alpha_4$ in $p$ implicitly. In particular,
when $\alpha_3=0$ the entropy in Eq.(\ref{BTZentropytor}) returns to the usual BTZ black hole entropy. It means that the N-Y term
$\alpha _{4}e^{a}\wedge T_{a}$  influences the BTZ metric and the exact form of entropy only when the CS term exists.

\section{Phase transition of the BTZ black hole}

It is well known that the heat capacity of BTZ black holes in GR is always positive. This point makes it different from its counterpart,
Kerr black holes in $2+1$ dimensions,  the heat capacity of which exists a discontinuous jump from negative one to positive one.
In the MB model the heat capacity at constant angular momentum $J$ of BTZ black holes can be expressed as
\be
C_J=\frac{\partial M}{\partial T}|_J
\ee
where $M, J, T$ are given in Eqs.(\ref{temp}) and (\ref{charge}).
Combing Eq.(\ref{temp}) with  Eq.(\ref{horizon}), one can derive a relation
\be
2\pi^2l^2T^2(1+\Delta)=M_0\Delta^2
\ee
From Eq.(\ref{charge}), one can express $M_0, J_0$ as functions of $M, J$, which are
\be
M_0=\frac{(1+\pi\alpha_3p)M+2\pi\alpha_3J/l^2}{A}, \quad J_0=\frac{(1+\pi\alpha_3p)J+2\pi\alpha_3M}{A}
\ee
where $A=(1+\pi\alpha_3p)^2-4\pi^2\alpha_3^2/l^2$ is a constant.

One can easily obtain the heat capacity at constant angular momentum
\be\label{hc}
C_J=\frac{4\pi r_+\Delta}{B(2-\Delta)-D(2+\Delta)(1-\Delta)/\sqrt{1-\Delta^2}}
\ee
where $B=\frac{1+\pi\alpha_3p}{A}$, $D=\frac{2\pi\alpha_3}{Al}$ are both constants.
Obviously when $\alpha_3=0$ it will return to the usual
heat capacity of BTZ black holes in GR, which is $C_{J_0}=\frac{4\pi r_+\Delta}{2-\Delta}$\cite{Cai3}. When $J_0=0$, namely non-rotating BTZ black holes,
$\Delta=1$, thus $C_J=C=4\pi r_+/B=4\pi l\sqrt{M_0}/B$. When $J_0=M_0l$, namely the extremal case, $\Delta=0$, thus $C_J=0$ ($B\neq D$) which corresponds to the vanishing
Hawking temperature.

Several comments are needed.
First, although when $\alpha_3=0$,  the conserved charges, entropy, the heat capacity of BTZ black holes in the MB model  all have the  expressions similar
to the ones of BTZ black holes in GR, this will not make the theory the same as the GR because of the different actions. Do not forget the $~l~$ is just
an effective one which is dependent on the parameters $\alpha_3, \alpha_4, \Lambda$. In fact they will be still different
so long as $\alpha_4$ is not equal to zero. Moreover, the existence of Nieh-Yan term
in the action remainds us the theory under consideration is a gravitational theory with torsion. Correspondingly the geometry on which the theory is based
 is Einstein-Cartan geometry. Second, the  heat capacity at constant angular momentum for the usual BTZ black hole in GR is always positive because $\Delta\in[0,1]$.
  But the one in the  MB model can be negative or positive, which means the existence of a second-order phase transition in the sense of Davies\cite{Davies},
  who first studied the heat capacity of Kerr-Newman black holes and claimed that a second-order phase transition will emerge when black holes get across
  the discontinuity point of heat capacity.

Below we will discuss the  heat capacity and analyze the critical phenomena of BTZ black holes in the MB model.
From Eq.(\ref{hc}), in order to have a divergence in $C_J$ the following condition must be satisfied
\be
 B(2-\Delta)-D(2+\Delta)(1-\Delta)/\sqrt{1-\Delta^2}=0
 \ee
It can be easily verified that $(2-\Delta)\geq (2+\Delta)(1-\Delta)/\sqrt{1-\Delta^2}$ always succeeds because of $0\leq\Delta\leq 1$,
thus $B\leq D$. This means that the conserved charge
$M$ can be negative.
Because of the  parameters $M_0, l, B, D$, the analytical solution for the heat capacity is a little problematic. Thus
we try to give some numerical solutions by assigning some values to the parameters. We can take $B, D, l$ or $\alpha_3, \alpha_4, \Lambda$
as the free parameters. Here we use the former one for simplicity. Firstly, we set $B=0.5, D=0.6$. The critical value of $\Delta_c$ can be
easily obtained, $\Delta_c= 0.748048$. According to  Eq.(\ref{horizon}), the corresponding critical value of $r_+$ is dependent on $M_0$ and $l$. We should express
$C_J$ as function of $r_+$ with parameters $M_0, l, B, D$.
\be
C_J=\frac{4\sqrt{2}\pi r^2_+/l\sqrt{M_0}}{\sqrt{2}B(3-2r^2_+/M_0l^2)r_+/l\sqrt{M_0}-\sqrt{2}D(1+2r^2_+/M_0l^2)\sqrt{1-r^2_+/M_0l^2}}
\ee
Obviously in order to have real roots, $1-r^2_+/M_0l^2\geq 0$ must be satisfied, namely $0\leq r_+\leq l\sqrt{M_0}$.

Numerical analysis shows that there are six roots for $r_+$, four imaginary roots, one negative real root
and one positive real root $r_e$ which is just the critical one. In table I we give the critical value of $r_+$ where the phase transition occurs.
According to the table it is evident that the value of $r_e$ depends on the choices of the parameters $M_0, l, B, D$. For fixed $B, D, l$ the greater $M_0$ gives
the greater $r_+$. Similarly for  fixed $B, D, M_0$ the bigger the values of $l$ are, the greater the $r_e$ is. Moreover the choices of $B, D$ have no significant influence on the values
of $r_+$. These values of $r_+$ in Table I all satisfy the condition $r_+\leq l\sqrt{M_0}$ obviously.

\begin{table}[htb]
\centering
\begin{tabular}{c c c c c c c}
\hline\hline
$B$ & $D$  & $M_0$ &$l$ & $r_e $   \\ [0.05ex]
\hline
0.5 & 0.6 & 1 & 1 & 0.934893 \\
0.5 & 0.6 & 0.5 & 1 & 0.661069 \\
0.5 & 0.6 & 2 & 1 & 1.32214 \\
0.5 & 0.6 &1 & 0.5  & 0.467446 \\
0.5& 0.6 & 1 & 2 & 1.86979 \\
0.4 & 0.6 & 1 & 1 &0.967353\\
0.2 & 0.6 & 1 & 1 & 0.993451 \\[0.05ex]
\hline
\end{tabular}\label{E1}
\caption{Numerical solutions for $r_+$ for given values of $M_0, l, B, D$ based on the conditions $B\leq D$ and $0 \leq \Delta\leq 1$.}
\label{tab}
\end{table}

Fig.1-3 show the relations between $C_J$ and $r_+$, which reveal that the phase transition occurs at the critical point $r_e$. In the figures one can  see
that except the transition point $r_e$ there exist another point $r_0$ where the heat capacity will change from positive one to negative one.   The heat capacity is positive when $r_+<r_0$ and $r_+>r_e$, while
it is negative for $r_0<r_+<r_e$. This character is a little like the critical behavior of Born-Infeld AdS black holes\cite{BIAdS}. Nevertheless, the point
$r_0$ is not the transition point because no discontinuity of heat capacity turns up.

\begin{figure}[tbp]
\centering
\includegraphics[angle=0,width=9cm,keepaspectratio]{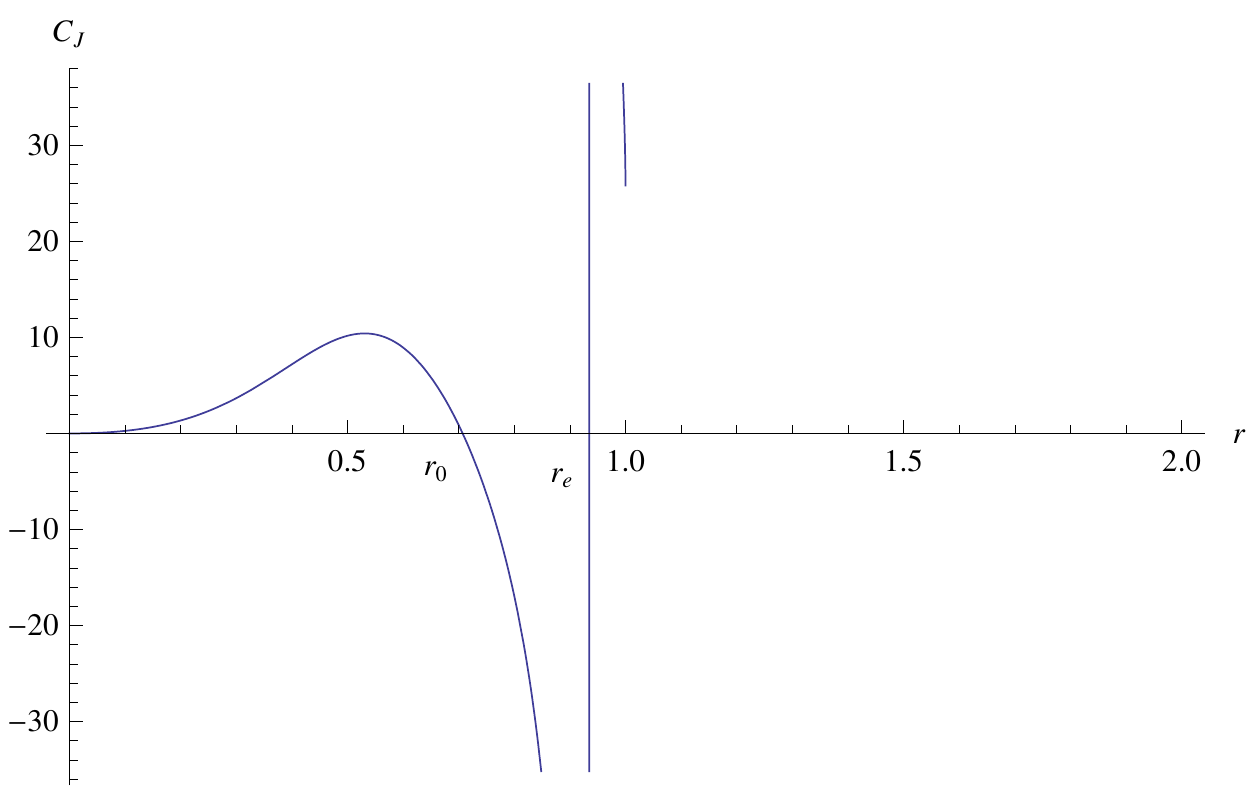}
\caption[]{\it Discontinuity of heat capacity $ C_{J} $ for BTZ black hole in the MB model at $r_{+}= 0.934893$ for $B=0.5$, $ D=0.6 $, $ M_0=1 $ and $ l=1 $}
\label{figure m}
\end{figure}
\begin{figure}[tbp]
\centering
\includegraphics[scale=0.5,keepaspectratio]{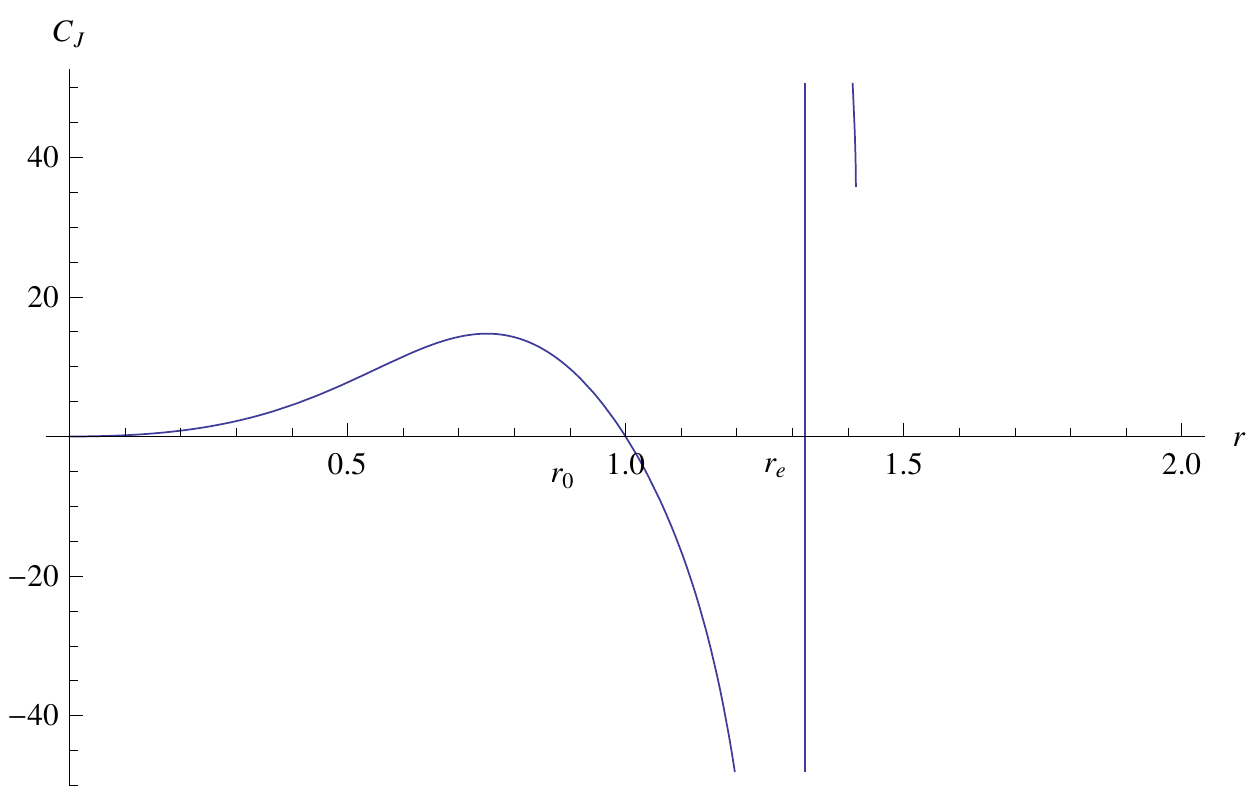}
\hfill
\includegraphics[scale=0.5,keepaspectratio]{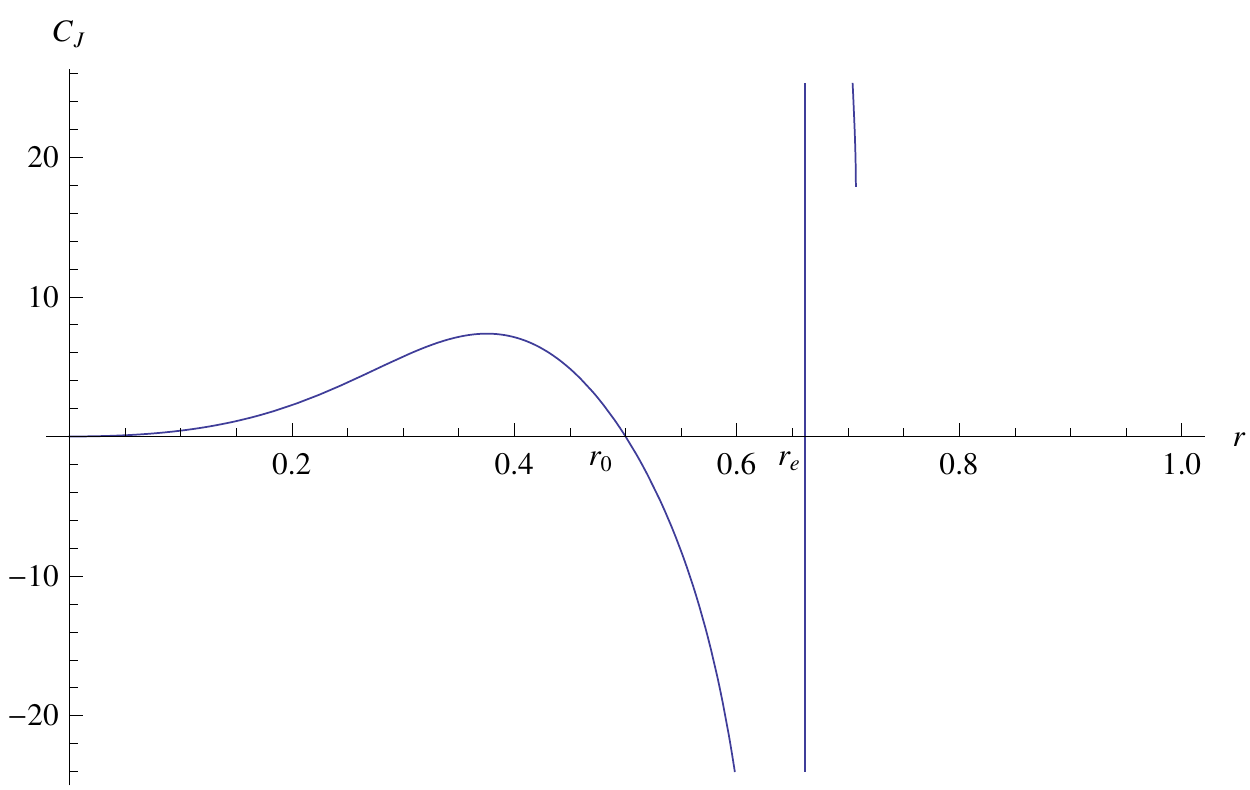}
\caption[]{\it Discontinuity of heat capacity $ C_{J} $ for BTZ black hole in the MB model at $r_{+}= 1.32214$  for
$B=0.5$, $ D=0.6 $, $ M_0=2 $, $ l=1 $,  and at $r_+=0.661069$  for
$B=0.5$, $ D=0.6 $, $ M_0=0.5 $, $ l=1 $}
\label{figure m12}
\end{figure}

\begin{figure}[tbp]
\centering
\includegraphics[scale=0.5,keepaspectratio]{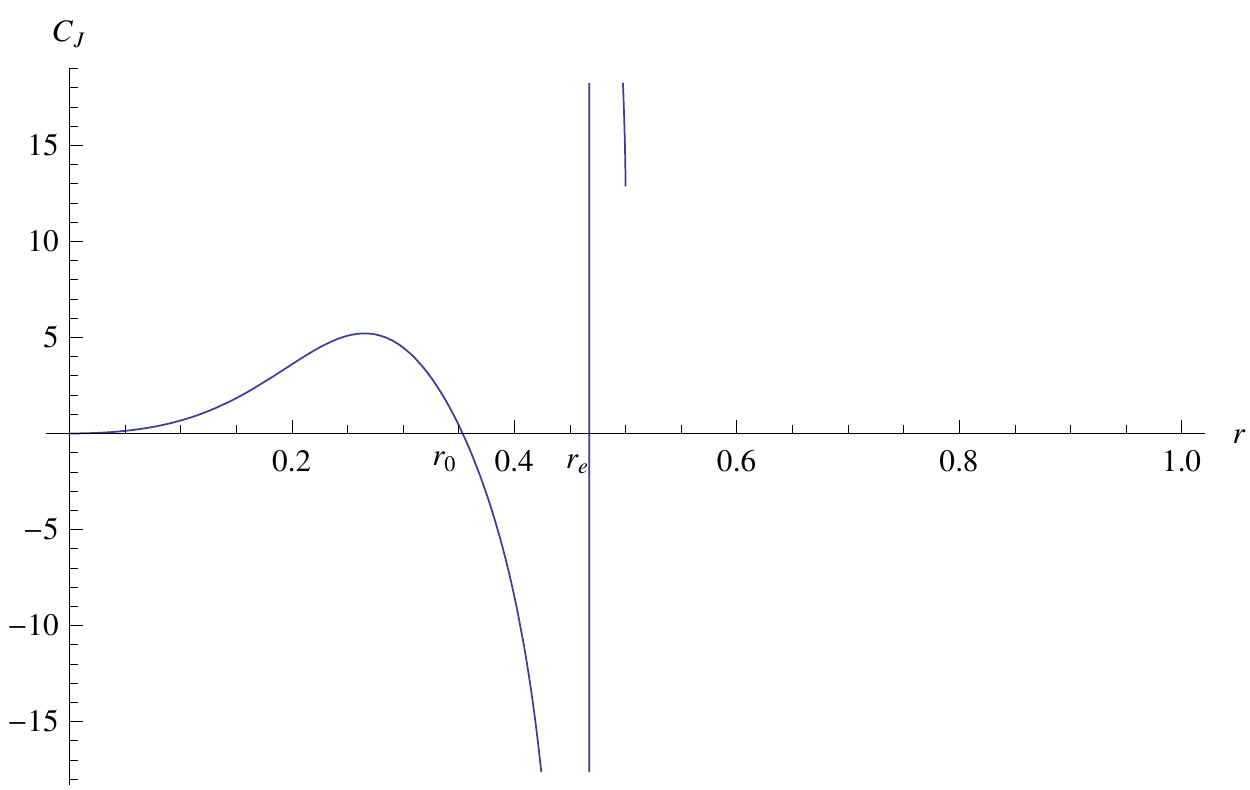}
\hfill
\includegraphics[scale=0.5,keepaspectratio]{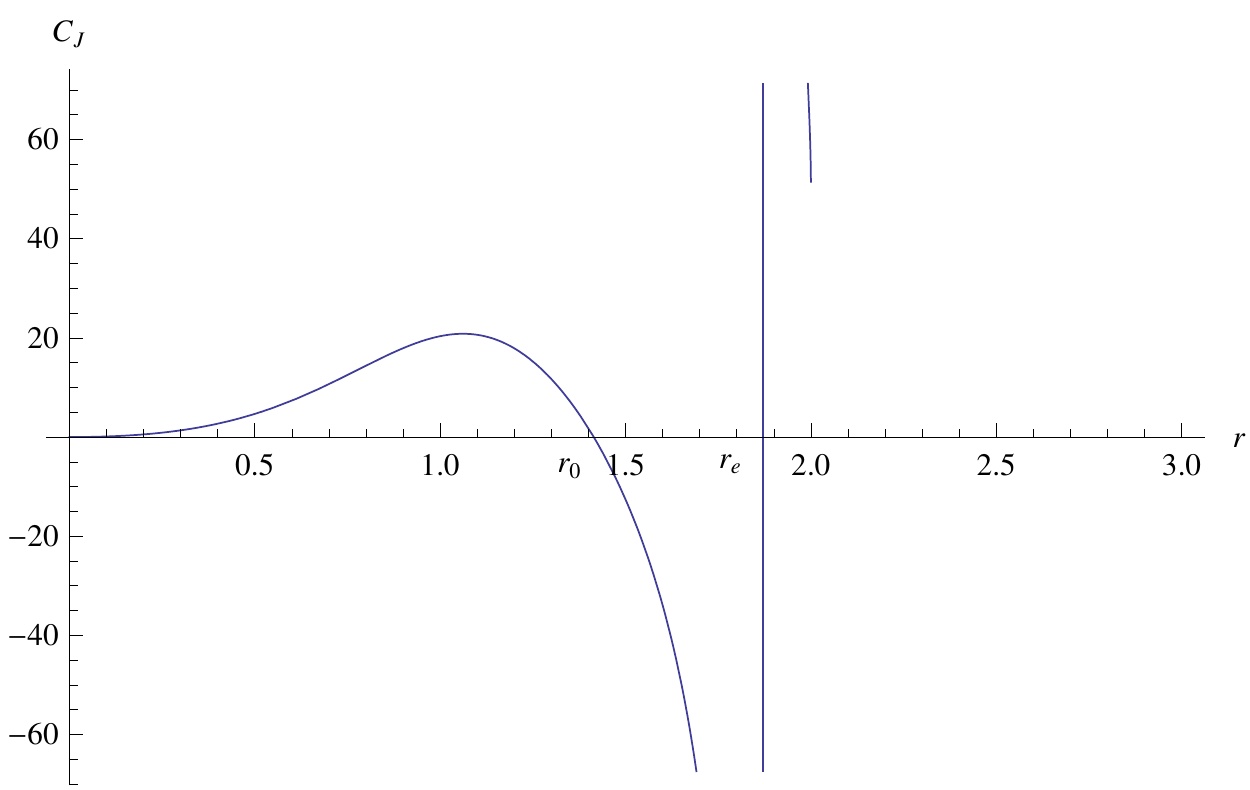}
\caption[]{\it  Discontinuity of heat capacity $ C_{J} $ for BTZ black hole in the MB model at $r_{+}= 0.467446$  for
$B=0.5$, $ D=0.6 $, $ M_0=1 $, $ l=0.5 $, and at $r_+=1.86979$ for
$B=0.5$, $ D=0.6 $, $ M_0=1 $, $ l=2 $}
\label{figure l23}
\end{figure}

Besides the phase transition in the sense of Davies above, there is another viewpoint that the phase transition will take place from the extremal to
nonextremal black holes\cite{Pavon1,Pavon2,Cai1,Cai2,Cai3}. According to the fluctuation theory of equilibrium thermodynamics, employing the Massieu function $\Phi$ to characterize
the equilibrium states of a thermodynamic system.
\be\label{fluc}
d\Phi=\sum_{i=1}^{n}X_idx_i
\ee
where the variables ${x_i}$ are called intrinsic variables and are determined by the thermodynamic environment. The ${X_i}$ are the conjugate
 variables of ${x_i}$\cite{Kaburaki1,Kaburaki2}. Below we will analyze the equilibrium fluctuations of BTZ black holes in the MB model
  in the microcanonical ensemble. In microcanonical ensemble nothing can be exchanged with the environment. The proper Massieu function is the entropy of system.
  For the BTZ black hole under consideration, it will be
  \be
  d\Phi=dS=\beta dM-\mu dJ
\ee
where $\beta=1/T$ and $\mu=\beta\Omega_H$. Comparing the above equation with Eq.(\ref{fluc}), the intrinsic variables are $\{M,J\}$ and the corresponding conjugate
variables are $\{\beta,-\mu\}$. One can easily derive the eigenvalues for fluctuation mode
\be
\lambda_M=\frac{\partial M}{\partial \beta}\mid_J=-T^2C_J, \quad \lambda_J=-\frac{\partial J}{\partial \mu}\mid_M=-TI_M
\ee
where $C_J$ is the heat capacity and
\be
I_M\equiv\beta\frac{\partial J}{\partial \mu}\mid_M=\left[\frac{B}{2r^2_+}-\frac{M_0\mu\Delta}{2\pi r^2_+}\frac{\partial r_+}{\partial J}\mid_M
-\frac{D\mu\Delta}{2\pi lr_+}-\frac{M_0\mu}{2\pi r_+}\frac{\partial\Delta}{\partial J}\mid_M\right]^{-1}
\ee
is the moment of inertia of BTZ black holes, where
\ba
\frac{\partial r_+}{\partial J}\mid_M&=&\frac{lD(1+\Delta)+M_0l^2\frac{\partial\Delta}{\partial J}\mid_M}{4r_+}, \nonumber\\
\frac{\partial\Delta}{\partial J}\mid_M&=&\frac{D}{l}\frac{1-\Delta^2}{\Delta M_0}-\frac{B(1-\Delta^2)}{\Delta J_0}
\ea
Some second moments can be easily obtained:
\ba
&&<\delta\beta\delta\beta>=-\frac{k_B\beta^2}{C_J}, \quad  <\delta\mu\delta\mu>=-\frac{k_B\beta}{I_M} \nonumber \\
&&<\delta\Omega_H\delta\Omega_H>=-k_B\left[\frac{T}{I_M}+\frac{\Omega^2_H}{C_J}\right], \quad <\delta\Omega_H\delta\beta>=\frac{k_B\beta\Omega_H}{C_J}
\ea
Thus for nonextremal black holes the second moments are all finite. when the extremal limit $\Delta=0$ is approached, because of $C_J=0, I_M=0$,
the second moments above will diverge. The divergence means the extremal BTZ black hole is a critical point. This conclusion is the same as the one for
the BTZ black hole in GR. In particular for the BTZ black hole in the MB model
the entropy is also a homogeneous function of $M, J$. Therefore the scaling laws at the critical point is expected to be still satisfied.

At last we mention the special case, $B=D$. According to the heat capacity (\ref{hc}), in the extremal case it is
\be
C_J=\lim\limits_{\Delta\rightarrow 0}\frac{4\pi r_+\Delta}{B(2-\Delta)-D(2+\Delta)(1-\Delta)/\sqrt{1-\Delta^2}}
\ee
If $B\neq D$ it will give the $C_J=0$. When the condition $B=D=finite ~ value$ is satisfied, it shows that $C_J\rightarrow \infty$. Thus in this case
the divergence point of heat capacity is just the extremal point. But unfortunately, $B=D$ cannot be satisfied. According to their definition, $B=D$ means
$A=0$. Thus $B=D=\infty$, which is irrelevant.  Therefore the two kinds of phase transition cannot be unified in one form.

\section{Entropy spectrum and the area spectrum of the BTZ black hole}

In what follows we firstly concentrate on analyzing non-rotating BTZ black hole.
When requiring $J_0=0$, the inner horizon $r_-$  vanishes. Thus
the angular velocity of BTZ black hole $\Omega_H=0$. The first law of
thermodynamics turns to be
\be
dM=TdS
\ee
Based on these conditions we calculate the area spectrum of BTZ black hole in the MB model according to QNMs.

Recently the Dirac quasinormal modes of non-rotating BTZ black hole with torsion in the MB model have been obtained\cite{Becar}, which are
\ba
\omega  &=&\frac{\xi }{l}-i\sqrt{M_0}\left( \frac{1}{2l}+\frac{2}{l}n+m+\frac{%
3p}{4}\right) ~, \no\\
\omega  &=&-\frac{\xi }{l}-i\sqrt{M_0}\left( \frac{3}{2l}+\frac{2}{l}n+m+\frac{%
3p}{4}\right) ~, \quad n=0,1,2 ...
\ea
where $\xi$ corresponds to the eigenvalue of Dirac operator.

According to Kunstatter\cite{Kunstatter}, there exists an adiabatic invariant
\be\label{int}
I=\int \frac{dE}{\omega(E)}
\ee
In general $dE$ in the nominator can be set equal to $dM_0$, but for the BTZ black hole with torsion
it does not hold any more.
From Eq.(\ref{charge})
\be
dE=dM=(1+ \pi\alpha_3p)dM_0
\ee
Fortunately although  the energy $E$ is not equal to the mass $M_0$, the first law of black hole thermodynamics still holds. Moreover in the
large $n$ limit, via Bohr-Sommerfeld quantization the adiabatic invariant  can be expressed as
\be
I=n\hbar
\ee
If considering the real part of the quasinormal modes as the fundamental vibrational frequency for the BTZ black hole with torsion.
one can obtain $\omega(E)\approx\omega_R=\pm\xi/l$. Thus
\be
I=\int \frac{(1+ \pi\alpha_3p)dM_0}{\omega_R}=\pm\frac{l(1+ \pi\alpha_3p)M_0}{\xi}+C
\ee
where $C$ is a integration constant. The mass spectrum is still equally spaced  with modified coefficient
\be
M_0=\pm\frac{\xi\hbar n}{l(1+ \pi\alpha_3p)}
\ee
The mass spacing is $\Delta M_0=\pm\frac{\xi\hbar}{l(1+ \pi\alpha_3p)}$. According to Eq.(\ref{horizon}) and $A=2\pi r_+$, the area spectrum is not equally
spaced.

In the view point of Maggiore, one can treat the perturbed black holes as a damped harmonic oscillator\cite{Maggiore}.
Thus the physically relevant vibrational frequency should be
\be
\omega(E)=\sqrt{\omega^2_R+\mid\omega^2_I\mid}
\ee
where $\omega_R$ and $\omega_I$ are respectively the real and imaginary parts of the quasinormal mode frequency.
In the large damping case, $\omega_R\ll\omega_I$, therefore
\be
\omega(E)\approx\mid\omega_I\mid
\ee
 In semiclassical limit
the characteristic frequency can be identified  with a transition between adjacent quasinormal frequency, namely the $\omega(E)$ in Eq.(\ref{int}) should be
replaced by
\be
\Delta\omega=(\mid\omega_I\mid)_{n}-(\mid\omega_I\mid)_{n-1}=2\sqrt{M_0}/l
\ee
Using the $\Delta\omega$ we can obtain the adiabatic invariant
\be
I=l(1+ \pi\alpha_3p)\sqrt{M_0}=n\hbar
\ee
or
\be
\sqrt{M_0}=\frac{n\hbar}{l(1+ \pi\alpha_3p)}
\ee
According to $r_+=l\sqrt{M_0}$ one can easily find that the horizon area is quantized and equally spaced as follows:
\be
A=2\pi r_+=\frac{2\pi n\hbar}{1+ \pi\alpha_3p}
\ee
For the non-rotating BTZ black hole in the MB model the black hole entropy is quantized and can be expressed as
\be
S=(1+\pi\alpha_3 p)4\pi r_+=4\pi\hbar n
\ee

Without using the QNMs, one can also calculate the entropy spectrum of black holes  via adiabatic invariance\cite{Vagenas}.
The adiabatic covariant action is
\be
I_{adia}=\oint p_i dq_i=2\pi nh
\ee
Employing the Hamiltonian $H$ of the system, which is the total energy of the black hole,
 and corresponding Hamiltonian's equations, for ordinary rotating BTZ black hole, the adiabatic covariant
action can be expressed as\cite{LWB,LHL}
\be\label{adia}
I_{adia}=\oint p_i dq_i=\hbar\int_{(0,0)}^{(M_0,J_0)}\frac{dM_0'-\Omega_HdJ_0'}{T}=\hbar S_0
\ee
For the BTZ black hole in the MB model the conserved charges are modified and the Hamiltonian of the system should be
$M$ but not $M_0$. According to Eq.(\ref{firstlaw}),the Eq.(\ref{adia}) still holds except changing $M_0, J_0, S_0$ into
$M, J, S$. Thus the entropy spectrum of the rotating BTZ black hole in the MB model is
\be
S=2\pi n, \quad n=1,2,3...
\ee
According to Eq.(\ref{BTZentropytor}), the area spectrum is not quantized because of the existence of $r_{-}$.
But for the non-rotating case the entropy spectrum and area spectrum are both quantized and equally spaced.

\section{Discussion and Conclusions}

In this paper we studied the phase transition of the BTZ black hole obtained in the MB topological gravitational model.
Although the gravitational action contains torsion, the BTZ solution is independent of torsion except the effective cosmological
constant refers to the coefficient $\alpha_4$ of the Nieh-Yan term. Because of the existence of  Chern-Simons term and Nieh-Yan term
the conserved charges for the BTZ black hole should be corrected. It is these corrections lead to the different thermodynamic properties and
critical behavior.

By calculating the heat capacity at constant angular momentum we find that it is not always positive any more, but
can change signs. Thus at the points where the heat capacity diverge, the phase transition happens.
This means that phase transition of black holes depends on the concrete theory of gravity. For the same BTZ black hole
 in different theories of gravity there are different critical behaviors.
Assigning some values to the free parameters
$B, D, l, M_0$ we obtain some phase transition points from a numerical analysis. From the figures one can see that
two points exist, where the heat capacity change signs. At the point $r_0$ the heat capacity change from positive one to negative one, but do not diverge.
While at the another point $r_e$ the heat capacity will diverge and change from negative one to positive one. Therefore the $r_e$ is the critical point where
the phase transition turns up.   Moreover we calculate some second moments of the thermodynamic quantities for the BTZ black hole in microcanonical ensemble.
It shows that they are finite in the  non-extremal case and diverge in the extremal limit of BTZ black hole. It means the phase transition happens from
the extremal to non-extremal BTZ black hole. This conclusion is the same as the one for the
ordinary BTZ black hole obtained in GR.

It should be noted that about the critical phenomena of the BTZ black hole we only studied the phase transition.
In fact the critical exponents are also very important. Whether the critical exponents are universal or depend on the details of the gravitational theory will
be considered in the next work. Furthermore, we will further study whether the behavior of the BTZ black hole obtained in the Chern-Simons  modified gravity
 or in the MB model is analogous to the Van der Waals fluid when treating the effective cosmological constant as a thermodynamic pressure.

In this paper we also discussed the area spectrum of the non-rotating BTZ black hole in the MB model according to quasinormal modes.
In the given QNM frequency no torsion exists. Thus this cannot reflect the influence of torsion on the area spectrum directly.
But the corrected mass $M$ will influence the quantized area. According to Maggiore's viewpoint one can still obtain the quantization of horizon area,
 however the quantized entropy of BTZ black hole is independent of the parameters $\alpha_3, p$. In addition we also simply discussed
 the method of adiabatic covariant action to obtain
the entropy spectrum. Because the conserved charges and the first law of black hole thermodynamics in the MB model are corrected, the result is different.
But the quantization of black hole entropy is not changed. This indicates the area spectrum and the entropy spectrum of black holes are not only dependent on the
spacetime metric, but also relevant to the theories of gravity.

\bigskip

\acknowledgments
This work is supported by NSFC under Grant
Nos.(11247261;11075098). M.-S Ma would like to thank Professor Chao-Guang Huang for useful discussion.

\end{document}